\documentclass[twocolumn]{aastex61}
\usepackage{hyperref}

\newcommand{\Msolar}{$M_{\odot}$}
\newcommand{\eMsolar}{M_{\odot}}

\newcommand{\myr}{\rm Myr}
\newcommand{\gyr}{\rm Gyr}
\newcommand{\dex}{\rm dex}

\newcommand{\ksf}{$k_{\rm SF}$}
\newcommand{\kof}{$k_{\rm OF}$}
\newcommand{\eksf}{k_{\rm SF}}
\newcommand{\ekof}{k_{\rm OF}}

\newcommand{\eMism}{M_{\rm ISM}}

\newcommand{\eMstar}{M_{\star}}

\newcommand{\eMhalo}{M_{\rm halo}}

\newcommand{\Msub}{$M_{\rm sub}$}
\newcommand{\eMsub}{M_{\rm sub}}

\newcommand{\Nsub}{$N_{\rm sub}$}
\newcommand{\eNsub}{N_{\rm sub}}

\newcommand{\eNccsn}{N_{\rm CCSN}}

\newcommand{\Nnsm}{$N_{\rm NSM}$}
\newcommand{\eNnsm}{N_{\rm NSM}}

\newcommand{\tnsm}{$t_{\rm NSM}$}
\newcommand{\etnsm}{t_{\rm NSM}}

\newcommand{\feh}{$[{\rm Fe/H}]$}
\newcommand{\efeh}{[{\rm Fe/H}]}

\newcommand{\bah}{$[{\rm Ba/H}]$}

\newcommand{\ecfe}{[{\rm C/Fe}]}

\newcommand{\mgfe}{$[{\rm Mg/Fe}]$}
\newcommand{\emgfe}{[{\rm Mg/Fe}]}

\newcommand{\eufe}{$[{\rm Eu/Fe}]$}
\newcommand{\eeufe}{[{\rm Eu/Fe}]}

\newcommand{\bafe}{$[{\rm Ba/Fe}]$}
\newcommand{\ebafe}{[{\rm Ba/Fe}]}

\newcommand{\srfe}{$[{\rm Sr/Fe}]$}
\newcommand{\esrfe}{[{\rm Sr/Fe}]}

\newcommand{\srba}{$[{\rm Sr/Ba}]$}
\newcommand{\esrba}{[{\rm Sr/Ba}]}

\newcommand{\rfe}{$[r{\rm /Fe}]$}
\newcommand{\erfe}{[r{\rm /Fe}]}

\newcommand{\rp}{$r$-process}
\newcommand{\rpe}{$r$-process element}
\newcommand{\rpes}{$r$-process elements}
\newcommand{\sfr}{{\rm SFR}}
\newcommand{\ofr}{{\rm OFR}}
\newcommand{\ba}{{\rm Ba}}
\newcommand{\eu}{{\rm Eu}}
\newcommand{\sr}{{\rm Sr}}
\newcommand{\mg}{{\rm Mg}}
\newcommand{\fe}{{\rm Fe}}

\usepackage{xcolor}
\newcommand{\SW}[1]{\color{black} #1}	

\usepackage{ulem} 

\shorttitle{Chemical Evolution of Subhalos and \lowercase{$r$}-process Elements}

\shortauthors{Ojima et al.}

\begin{document}

\title{Stochastic Chemical Evolution of Galactic Subhalos and the Origin of \lowercase{$r$}-process Elements}

\author{Takuya Ojima}
\affil{Department of Material Science, International Christian University, 3-10-2 Osawa, Mitaka, Tokyo 181-8585, Japan}

\author{Yuhri Ishimaru}
\affil{Department of Material Science, International Christian University, 3-10-2 Osawa, Mitaka, Tokyo 181-8585, Japan}

\author{Shinya Wanajo}
\affil{Department of Engineering and Applied Sciences,
Sophia University, Chiyoda-ku, Tokyo 102-8554, Japan;
shinya.wanajo@sophia.ac.jp}
\affil{iTHEMS Research Group, RIKEN, Wako, Saitama 351-0198, Japan}

\author{Nikos Prantzos}
\affil{Institut d'Astrophysique de Paris, UMR7095 CNRS, Univ. P. \& M. Curie, 98bis Bd. Arago, F-75104 Paris, France}

\author{Patrik Fran\c{c}ois}
\affil{GEPI, Observatoire de Paris, PSL Research University, CNRS, 61 Avenue de l'Observatoire, F-75014 Paris, France}
\affil{Universit\'{e} de Picardie Jules Verne, 33 rue St Leu, Amiens, France}


\begin{abstract}

Mergers of compact binaries (of a neutron star and another neutron star or a black hole, NSMs) are suggested to be the promising astrophysical site of the \rp. 
While the average coalescence timescale of NSMs appears to be $\gtrsim 100 \, \myr$, most of previous chemical evolution models  indicate that the observed \SW{early appearance and large} dispersion of \rfe\ in  Galactic halo stars at $\efeh \lesssim -2.5$  favors shorter coalescence times of 1--10 \myr. We argue that this is not the case for the models assuming the formation of the Galactic halo from clustering of subhalos with different star formation histories as suggested by Ishimaru et al. We present a stochastic chemical evolution model of the subhalos, in which the site of the \rp\ is assumed to be mainly  NSMs with a coalescence timescale of $100 \, \myr$. \SW{In view of the scarcity of NSMs, their} occurrence in each subhalo is computed with a Monte Carlo method. Our results show that the less massive subhalos evolve at lower metallicities and generate highly \rp-enhanced stars. An assembly of these subhalos leaves behind the large star-to-star scatters of \rfe\ in the Galactic halo as observed. However, the observed scatters of [Sr/Ba] at low metallicities indicate the presence of an additional site that partially contributes to the enrichment of light neutron-capture elements such as Sr. The high enhancements of \rfe\ at low metallicities found in our low-mass subhalo models also \SW{qualitatively reproduce} the abundance signatures of the stars in the recently discovered ultra-faint dwarf galaxy Reticulum II. Therefore, our results suggest NSMs as the dominant sources of \rpes\ \SW{in the Galactic halo}.
%
%
\end{abstract}




\section{Introduction}

Most of the elements with atomic numbers greater than $Z = 30$--40 are produced through neutron-capture processes, and about half of such heavy elements originate from the rapid neutron-capture process (\rp). However, the astrophysical site of the \rp\ has not been specified, which remains a long-standing problem in nuclear astrophysics (see, e.g., \citealt{Thielemann+17} for a recent review).

Stellar abundances of Galactic halo stars serve as the fossils of the early history of the Galaxy, providing us with important clues to the Galactic formation and early chemical evolution. Metal-poor stars, those with metallicities of \feh%
	\footnote{$[{\rm A} / {\rm B}] \equiv \log_{10}(X_{\rm A} / X_{\rm B}) - \log_{10}(X_{\rm A, \odot} / X_{\rm B, \odot})$, where $X_{\rm A}$ and $X_{\rm B}$ are the mass fractions of elements $\rm A$ and $\rm B$.}
 $\lesssim -2.5$, are thought to be among the oldest objects in the Galaxy, which presumably have been formed in the first few hundred Myr of its life. Spectroscopic observations of metal-poor halo stars show a large star-to-star scatter of about 2.5~dex in the abundances of Eu (as representative of \rp\ elements) with respect to Fe, \eufe\  (e.g., \citealt{Honda+04, Francois+07, SAGA, Sneden+08}). In particular, several metal-poor stars such as CS~22892-052 \citep{Sneden+03} and CS~310812-001 \citep{Siqueira+13} show extremely high ratios of $\eeufe = 1.6$--1.7. These unique abundance signatures indicate that \SW{Eu does} not share common \SW{astrophysical conditions} with $\alpha$ or iron-group elements.
 
It is also well known that such \rp-enhanced stars, which account for about 10\% of all metal-poor stars, exhibit fairly robust abundance distributions that agree with the solar system \rp\ pattern. On the one hand, the excellent agreement for the heavier side ($Z > 50$; \citealt{Sneden+08}) indicates the presence of the single robust ``main" \rp\footnote{Hereafter, we indicate the main \rp\ by the ``\rp", which produces all of the \rp\ elements with a solar \rp-like pattern but with a smaller content of $Z < 50$ elements.} site. On the other hand, the less remarkable agreement for the lighter side ($Z < 50$; \citealt{Siqueira+14}) as well as the bulk of (\rp-deficient) stars showing higher [Sr/Ba] than the solar \rp\ ratio \citep{McWilliam98, Johnson+02} implies the presence of another ``weak" \rp\ that produces only light neutron-capture elements \citep{Wanajo+06}. In fact, several metal-poor stars showing a descending trend of neutron-capture elements toward the heavier side have been identified \citep{Honda+06, Honda+07, Aokim+17}, which may reflect such a process.

The major candidates suggested as the \rp\ site include core-collapse supernovae (CCSNe; e.g., \citealt{Burbidge+57, Hillebrandt+84, Woosley+94}) and binary mergers of a neutron star and another neutron star or a black hole (NSMs\footnote{In this paper, we generally indicate ``neutron star--neutron star mergers" by ``NSMs", although similar conclusions may be obtained for neutron star--black hole mergers.}; e.g., \citealt{Lattimer+74, Symbalisty+82, Eichler+89, Meyer89, Freiburghaus+99, Goriely+11, Korobkin+12, Wanajo+14}). Inhomogeneous chemical evolution models \citep{Ishimaru+99, Argast+00, Tsujimoto+00, Ishimaru+04, Cescutti08}, which account for incomplete mixing of nucleosynthetic ejecta in the interstellar medium (ISM), have demonstrated that the observed dispersion of \rfe\ ratios (where $r$ indicates an \rpe) in Galactic halo stars can be explained if CCSNe from a limited initial stellar mass range are the sources of \rp\ elements. 
However, recent nucleosynthesis studies show difficulties in producing heavy \rp \ elements  $(Z \gtrsim 50)$ in the physical conditions relevant to CCSNe \citep{Wanajo+11, Wanajo+18, Wanajo13, Bliss+18}, which can be at best the sources of light neutron-capture elements made by a weak \rp. Effects of a strong magnetic field also have been discussed \citep{Thompson03, Suzuki+05, Winteler+12, Nishimura+15}, although their roles on the \rp\ are under debate \citep{Nishimura+17, Thompson+17, Moesta+17}.

By contrast, recent nucleosynthesis calculations based on the hydrodynamical simulations of NSMs reasonably reproduce the solar $r$-process abundance curve (\citealt{Wanajo+14, Goriely+15, Radice+16, Wu+16}). The discovery of an electromagnetic emission \citep[kilonova;][]{Li+98, Metzger+10} associated with the gravitational-wave source GW170817 \citep{Abbott+17} also supports NSMs as the site of the \rp\ in the universe. In fact, the inferred amount of the \rp\ material ejected from this event, about 0.03--$0.05\, M_\odot$ \citep[e.g.,][]{Pian+17}, appears to be sufficient to account for the total mass of \rp\ elements in the Galaxy, provided that GW170817 is representative of NSM events.

However, binary population synthesis models (e.g., \citealt{Dominik+12}) as well as observations of binary neutron stars \citep[e.g.,][]{Beniamini+16} suggest an average NSM coalescence timescale $\langle \etnsm \rangle \gtrsim 100 \, \myr$, which appears too long to allow for the observed appearance of, e.g., Eu at  metallicities as low as $\efeh\sim -2.5$ on the basis of one-zone chemical evolution models \citep{Argast+04}. Such models suggest $\etnsm = 1-10 \, \myr$ to reproduce the observed early evolution of \rfe\ \citep{Argast+04, DeDonder+04, Matteucci+14, Tsujimoto+14, Cescutti+15, Wehmeyer+15}. Previous chemical evolution models based on the hierarchical structure formation of the Galactic halo also favor short NSM coalescence timescales (\citealt{Komiya+14, vandeVoort+15}).  
In addition, it is argued that the inferred  low event rate ($0.4-77.4 \, \myr^{-1}$; \citealt{Dominik+12}) of NSMs causes too large \rfe\ dispersions at higher metallicity to be compatible with observations \citep{Qian00, Argast+04}.

Chemical evolution studies assuming multiple $r$-process sites such as NSMs and magnetorotationally driven CCSNe attempt to explain the observed \rfe\ evolution of the metal-poor stars (\citealt{Wehmeyer+15, Shibagaki+16}). 
However, the uniqueness of the abundance patterns in \rp-enhanced stars apparently disfavors multiple \rp\ sites with different abundance distributions \SW{\citep[e.g.,][]{Sneden+08}} considered in these studies.

\citet{Prantzos+06} suggested that the observed appearance of Eu at low metallicity as well as the large dispersion of [Eu/Fe] could be naturally explained if the Galactic halo was formed from merging  subhalos with different star formation histories and if the production sites of Fe and \rp\ elements evolved on different timescales. \citet[][hereafter IWP15]{Ishimaru+15} have first explored this idea using a semianalytical model of merging subhalos, each of them evolving on a different timescale---depending on its mass---in a homogeneous way (i.e., the gas is assumed to be well mixed within each subhalo). 
According to IWP15, NSMs start occurring at the metallicity $\feh \lesssim -3$ and contributing to the enrichment of Eu even with the coalescence timescale of 100~Myr if less massive subhalos evolve with lower star formation efficiency. Similar results can be found in recent semianalytic \citep{Komiya+16} and chemodynamical \citep{Shen+15, Hirai+15, Hirai+17} studies \SW{(see also \citealt{Cote+17} for a comparison of several chemical evolution studies mentioned above with their own model)}. 

IWP15 also show that various star formation efficiencies make a difference in the numbers of cumulative NSMs occurring in subhalos. In fact, the number of cumulative average NSMs for the lightest subhalo with a stellar mass of $10^4\, M_\odot$ (similar to that of an ultra-faint dwarf galaxy, UFD) predicted in IWP15 is $\sim 0.1$, implying that only one 1 of 10 small subhalos experiences an NSM event. They suggested that a single NSM occurring in the least massive systems would lead to a very high \rfe\ of the inter stellar medium. The recently discovered UFD, Reticulum II \citep[Ret~II;][]{Ji+16_Nature, Roederer+16, Ji+16_Sr, Ji+18}, could be such an example as anticipated by IWP15, in which seven (out of nine) stars exhibited high \rfe\ ratios comparable to those in the most \rp-enhanced Galactic halo stars.

In this paper, we extend the study of IWP15 to explain the presence of such \rp-enhanced stars and the scatter of \rfe\ ratios in the Galactic halo as well as in Ret~II. We also aim to examine the chemical evolution of Sr to test if our model is compatible with those of light neutron-capture elements. For this purpose, we construct a chemical evolution model, in which individual subhalos stochastically experience NSM events. In section \ref{sec:model}, the concept and setup of our model are presented in detail. In section \ref{sec:results}, we discuss the enrichment histories of Ba and Eu as representative of \rp\ elements. We also compare our results with the observations of a light neutron-capture element Sr (section~\ref{sec:strontium}) as well as of the \rp-enhanced stars in Ret II (section~\ref{sec:ufd}). Finally, we summarize and conclude our work in section \ref{sec:summary}.


\section{Chemical Evolution Model}
\label{sec:model}

Based on the hierarchical structure formation scenario, our Galactic halo is formed from subhalos with various masses.  Depending on their masses, the subhalos are expected to have different star formation histories, and the sum of the subhalos weighted by their mass function is assumed to become the Galactic halo. Within each subhalo, the occurrence of NSMs is treated stochastically (see Sec. 2.3).


\subsection{Chemical Evolution of Different Mass Subhalos}

The star formation rate $(\sfr)$ of a subhalo\SW{, $\psi(t)$,} is assumed to be proportional to its interstellar medium (ISM) mass, $\eMism(t)$,
	\begin{equation}
	\psi(t) = \eksf \, \eMism(t) ,
	\label{eq:sfr}
	\end{equation}
where \ksf\ is \SW{a time-independent} star formation efficiency. All subhalos are expected to suffer gas outflow because of their weak gravitational binding (as a result of, e.g., gas heating, ram pressure, and tidal stripping)  in which the interstellar material is efficiently mixed  (i.e., spatially homogeneous). The gas outflow rate $(\ofr)$\SW{, $\varphi(t)$,} is assumed to be proportional to the $\sfr$,
	\begin{equation}
	\varphi(t) = \eta \, \psi(t) ,
	\label{eq:ofr}
	\end{equation}
where $\eta$ \SW{is a time-independent coefficient}. 
We introduce \SW{a time-independent outflow efficiency,} $\ekof \equiv \eta \, \eksf$; i.e., $\varphi(t) = \ekof \, \eMism(t)$.

For the stellar initial mass function (IMF), we adopt Kroupa's IMF \citep{Kroupa02} within the mass range of 0.05--$1 \, \eMsolar$ \SW{and the slope of $-2.7$ is taken for the range of 1--$60 \, \eMsolar$} \citep[see also][]{Kubryk+15}. \SW{Massive stars of 10--$60 \, \eMsolar$ are the progenitors of CCSNe in this study.}

Although it is nearly impossible to directly observe ancient subhalos nowadays, they might be very close in nature to the local  dwarf spheroidal galaxies that we observe now. The observed mass-metallicity relation of those galaxies by \citet{Kirby+13} shows a clear correlation between the stellar masses and the mean metallicities of the local dwarf galaxies, regardless of their morphologies. Therefore, we assume that subhalos also have the same correlation. This mass-metallicity relation scales as $\langle\efeh\rangle \propto \log{\eMstar}^{0.3}$ within the mass range $10^3 - 10^{8} \, \eMsolar$, where $\langle\efeh\rangle$ and $\eMstar$ are the mean metallicity and the stellar mass of a dwarf galaxy, respectively. Here, we assume that the mean metallicity is equal to the peak of a given metallicity distribution. This leads to $\eta \propto {\eMstar}^{-0.3}$ (\citealt{Prantzos08}; IWP15). Thus, we adopt
	\begin{equation}
	\eta(\eMsub) = \eta_{8} \, \left( \frac{\eMsub}{10^8 \, \eMsolar} \right)^{-0.3} , \label{eq:eta}
	\end{equation}
where \Msub\ is the stellar mass of a subhalo and  $\eta_8$ corresponds to the value for a subhalo with the mass $10^8 \, \eMsolar$, which is one of the most massive subhalos assumed in our model. The values $\eta_8$ and $10^8 \, \eMsolar$ are chosen so that the subhalos with the final stellar mass $10^8 \, \eMsolar$ have a metallicity reaching $\efeh \sim -1$ at the final time $t = 2 \, \gyr$, which is realized with $\eksf = 0.20 \, \gyr^{-1}$ and $\ekof = 1.0 \, \gyr^{-1}$; i.e., $\eta_8 = 5.0$. Since Galactic halo stars show no trace of type~Ia supernovae, we do not include their contribution to Fe yields in our model. 

For each value of $\eta$, we consider two extreme cases in order to determine  \ksf\ and \kof\ (as in \SW{Table~1} in IWP15). For case 1, \kof\ is assumed to be constant at $\ekof = 1.0 \, \gyr^{-1}$ while $\eksf \propto {\eMsub}^{+0.3}$. On the other hand, for case 2, \ksf\ is assumed to be constant, $\eksf = 0.20 \, \gyr^{-1}$, while $\ekof \propto {\eMsub}^{-0.3}$. The $\sfr$ $\psi(\eMsub, t)$ is then obtained as a function of time and the stellar mass of a subhalo.

Element yields of CCSN are taken from \citet{Nomoto+13} \SW{with a linear interpolation between 10 and $40\, M_\odot$ and those at $40\, M_\odot$ in the range 40--$60\, M_\odot$}. We do not consider the enrichment from the $s$-process, because  the s-component of the heavy elements stems mainly from the AGB phase of low-mass stars and is not expected to contribute to metal-poor stars. Since we focus mainly on metal-poor stars, we also treat $\ba$ (an $s$-process-dominant element in the solar system) as an \rpe, a concept that was first suggested by \cite{Truran81}\SW{; see, however, \citet{Prantzos+18} for the case of rotating massive stars, which may have a non-negligible contribution to $s$-element production even at low metallicities}.  

The ISM is assumed to be well mixed; thus, enrichment by CCSNe is calculated in a continuous way by a simple chemical evolution model with gas outflow. Enrichment of \rpes\ is computed in a stochastic way only when NSMs occur.


\subsection{Subhalo Mass Distribution}

Cosmological simulations based on the hierarchical structure formation scenario predict the dark matter function of subhalos, which is proportional to the inverse square of the subhalo dark mass \citep{Diemand+07}, i.e., proportional to the inverse square of the baryonic mass. Taking into account the effective yield from the mass-metallicity relation, which suggests outflow, the final stellar mass is smaller than the initial baryonic mass. As a result, the subhalo stellar mass function becomes $\Phi(\eMsub)$ $\equiv$ $d\eNsub$/$d\eMsub$ $\propto$ ${\eMsub}^{-1.7}$ (\citealt{Prantzos08}; IWP15).

We consider the subhalos with stellar masses ranging from $10^4 \, \eMsolar$ to $2 \times 10^8 \, \eMsolar$. The lowest subhalo mass that we consider is comparable to the smallest local dwarf galaxies \citep{Kirby+13}\footnote{Note that the observed lowest stellar mass of the local dwarf galaxies from \citet{Kirby+13} is of the order of $10^3 \, \eMsolar$ (Segue 2); however, \SW{we set the lowest mass of subhalos be that of intermediate-mass ultra-faint dwarfs, $10^{4}\, M_\odot$, because the mass-metallicity relation scales well in the range $10^{3.5} < M_*/M_\odot < 10^9$ \citep[Figure~9 in][]{Kirby+13}.}}. The highest mass is set to half of the mass of the Galactic halo, $\eMhalo = 4\times10^8\, \eMsolar$ \citep{Bell+08}. The subhalo stellar mass function is normalized as follows:
	\begin{equation}
	\eMhalo = \int_{10^4\,\eMsolar}^{2\times10^8\,\eMsolar}
    				{ \eMsub \, \Phi(\eMsub) \, d\eMsub } .
	\end{equation}
By definition, the total number of subhalos within the mass range between $M_1$ and $M_2$ is given by
	\begin{equation}
	\eNsub(M_1, M_2) = \int_{M_1}^{M_2}
    						{ \Phi(\eMsub) \, d\eMsub } .
	\end{equation}
The third column of Table~1 shows the numbers of subhalos between $M_1$ and $M_2$ (first and second columns, respectively) according to Equation~(5). Note that $N_\mathrm{sub}$ is the same for both cases~1 and 2.
%
%
	\begin{deluxetable*}{cccccccc}
	\tablecolumns{8}
	\tablewidth{0pc}
	\tablecaption{Numbers of Subhalos and NSMs for Case~1}
	\tablehead{
		\colhead{$M_1$} &
		\colhead{$M_2$} &
		\colhead{\Nsub}  &
		\colhead{\Nnsm} &
		\colhead{\Nnsm} &
		\colhead{\Nnsm} &
        \colhead{$N_{\rm NSM,min}$} &
        \colhead{$N_{\rm NSM,max}$} \\
			\colhead{$(\eMsolar)$} &
			\colhead{$(\eMsolar)$} &
			\colhead{} &
			\colhead{(mean)} &
			\colhead{(min.)} &
			\colhead{(max.)} &
            \colhead{$/\eNccsn$} &
            \colhead{$/\eNccsn$}
		}
	\startdata
	$10^4$ & $10^5$ & $741$ & $0.0702$ & $0$ & $2$ & 
    						$0$ & $1.45 \times 10^{-3}$ \\
	$10^5$ & $10^6$ & $147$ & $0.918$ & $0$ & $5$ & 
    						$0$ & $3.63 \times 10^{-4}$ \\
	$10^6$ & $10^7$ & $29$ & $10.3$ & $2$ & $38$ & 
    						$1.45 \times 10^{-5}$ & $2.75 \times 10^{-4}$ \\
	$10^7$ & $10^8$ & $6$ & $102$ & $30$ & $229$ & 
    						$4.90 \times 10^{-5}$ & $3.74 \times 10^{-4}$ \\
	$10^8$ & $2\times10^8$ &
					  $1$ & $392$ & $392$ & $392$ & 
							$9.85 \times 10^{-4}$ & $9.85 \times 10^{-4}$ \\
	\enddata
	\label{tab:calcNSM}
	\end{deluxetable*}


\subsection{Enrichment by NSMs}

The suggested typical coalescence timescale of NSM in the literature is $\gtrsim 100 \, \myr$ (e.g., \citealt{Belczynski+08, Dominik+12}). Although these studies also predict short-lived mergers ($\lesssim 1 \, \myr$), the fraction of such binaries is estimated to be less than several percent \citep{Dominik+12}. In our model, we adopt the bimodal NSM coalescence timescales from IWP15, where the proportions of long-lived (100~Myr) and short-lived (1~Myr) NSMs are set to $95\%$ and $5\%$, respectively. The average frequency of NSM to CCSN is assumed to be $\langle f_{\rm NSM} / f_{\rm CCSN} \rangle = 1/1000$ according to the population synthesis model by \citet{Dominik+12}.

The number of CCSN events occurring within a given time interval $\Delta t$ is
	\begin{eqnarray}
	&\ & \Delta \eNccsn(\eMsub, t) \nonumber \label{eq:Nccsn} \\
	&=& \int_{t}^{t+\Delta t}{\!\!\!\!\! dt}
		\int_{10 \, \eMsolar}^{60 \, \eMsolar}
        		{ \psi(\eMsub, t-\tau_m) \, \phi(m) \, dm } \, ,
	\end{eqnarray}
where $\tau_m$ is the lifetime of the star with initial mass $m$ \SW{adopted from \citet{Schaller+92}}. The number of NSM events occurring within the time interval $\Delta t$ is then
	\begin{eqnarray}
	\Delta \eNnsm(\eMsub, t)
	= \frac{f_{\rm NSM}}{f_{\rm CCSN}} \, 
    \Delta \eNccsn(\eMsub, t-\etnsm), \label{eq:Nnsm}
	\end{eqnarray}
where \tnsm\ is the NSM coalescence timescale. Equations (\ref{eq:Nccsn}) and (\ref{eq:Nnsm}) account for the \textit{average} numbers of CCSNe and NSMs, respectively, occurring in a subhalo with mass \Msub\ in the corresponding time interval.

In our model, we first calculate the average number of NSMs occurring in each time interval \SW{(taken to be a few percent of $t$)} for a subhalo of $M_\mathrm{sub}$. The total average number of NSMs occurring in a group of subhalos with masses between $M_1$ and $M_2$ is then obtained as $\Delta \eNnsm \eNsub$ in the corresponding time interval. Using a Monte Carlo method, we randomly choose the subhalos in which NSMs occurred. In Table~1 the resulting average, minimum, and maximum numbers of NSMs (for case~1) in the subhalos of masses between $M_1$ and $M_2$ after 2~Gyr are presented in the 4th, 5th, and 6th columns, respectively.

The yield of Eu ejected by a single NSM event is assumed to be $y_\mathrm{Eu} = 4 \times 10^{-5}\, M_\odot$ so that the observed average [Eu/Fe] values are reproduced (Figures~4). This value is about a factor of 2 greater than that obtained by the nucleosynthesis calculation in \citet{Wanajo+14}, $\sim 2 \times 10^{-5}\, M_\odot$, which is, however, dependent on the still uncertain total ejecta mass from an NSM \citep[e.g.,][]{Shibata+17}. Besides $\eu$, we also calculate the chemical evolution of $\ba$ as well, its yield being $y_\mathrm{Ba} = 8\,  y_\mathrm{Eu}$, which matches the solar \rp\ abundance ratio \citep{Burris+00}. The parameters concerning NSMs are summarized in Table~2\footnote{\SW{Our parameter settings give the number of NSMs and iron mass per stellar mass, which are IMF-averaged over the entire mass range 0.05--$60\, M_\odot$, of $3.2\times 10^{-6}\, (M_\odot^{-1})$ and $2.8\times 10^{-4}$, respectively. These values are within the range of those from other recent chemical evolution models (Figure. 1 in \citet{Cote+17}. The normalized number of NSMs per stellar mass according to Eq.~(1) in \citet{Cote+17} is $1.5\times 10^{-5}\, (M_\odot^{-1})$, which also falls within the range of other models (their Figure~5).}}.
%
%
	\begin{deluxetable}{lccccr}
	\tablecolumns{6}
	\tablewidth{0pc}
	\tablecaption{NSM-related Model Parameters}
	\tablehead{
		\colhead{Type} & 
		\colhead{\tnsm} &
		\colhead{Prop.} &
		\colhead{$f_{\rm NSM} / f_{\rm CCSN}$} &
		\colhead{$y_{\eu}$} &
		\colhead{$y_{\ba}$} \\
			\colhead{} &
			\colhead{$(\myr)$} &
			\colhead{$(\%)$} &
			\colhead{} &
			\colhead{(\Msolar)} &
			\colhead{$(\eMsolar)$}
		}
	\startdata
	Short-lived & $1.00$ & $5$ & $5.00 \times 10^{-5}$ & $4\times10^{-5}$ & $8 \, y_{\eu}$ \\
	Long-lived & $100$ & $95$ & $9.50 \times 10^{-4}$ & $4\times10^{-5}$ & $8 \, y_{\eu}$ \\
	\enddata
	\label{tab:paramNSM}
	\end{deluxetable}


\section{Chemical evolution of \lowercase{$r$}-process elements}

\label{sec:results}


\subsection{Chemical Evolution of individual Subhalos}
\label{subsec:chemev}

\begin{figure*}
	\begin{center}
	\includegraphics[width=0.8\textwidth]{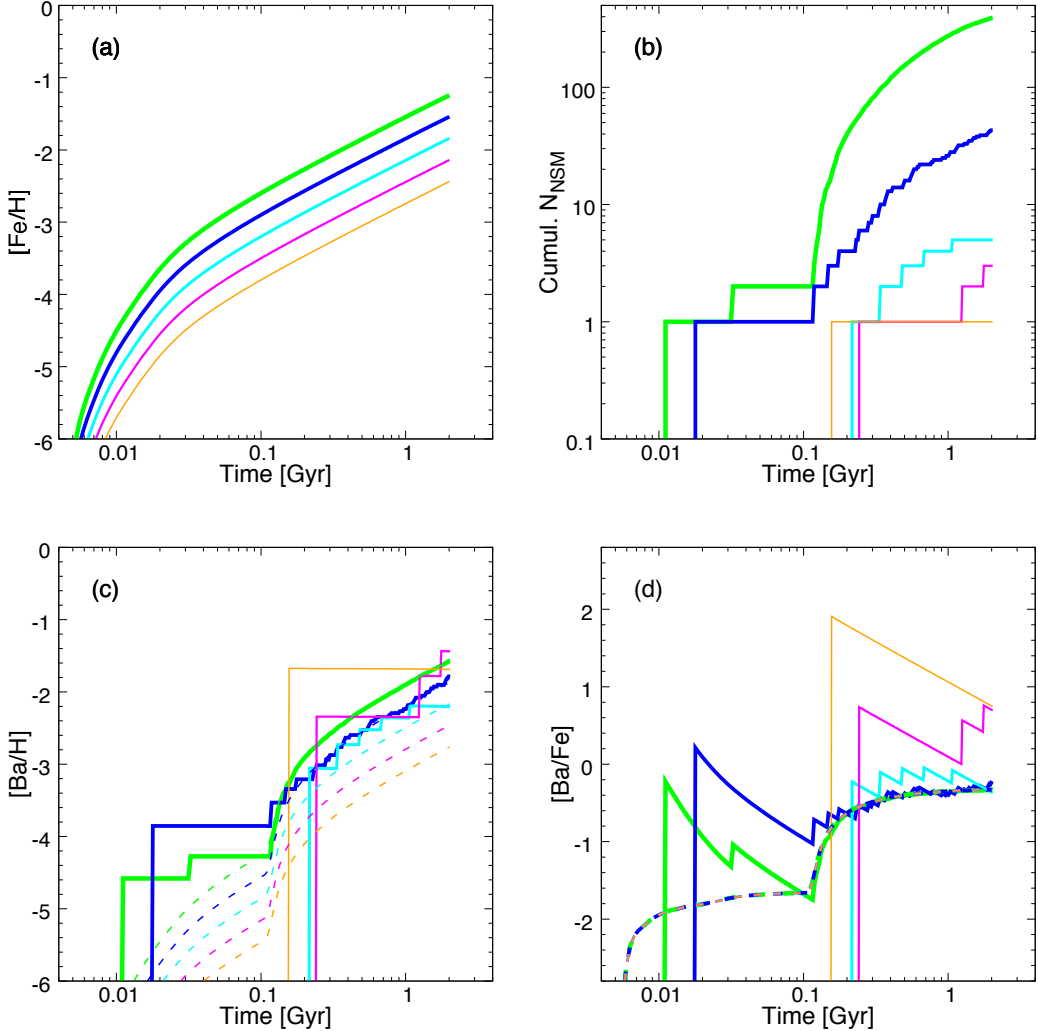}
	\caption{Chemical evolutions (for case~1) of five selected subhalos with stellar masses of $10^4$ (orange), $10^5$ (magenta), $10^6$ (cyan), $10^7$ (blue), and $10^8 \, \eMsolar$ (green) from the thinnest to thickest curves. (a) Age-metallicity relation. (b) Time evolution of the cumulative number of NSMs. (c) Time evolution of \bah. (d) Time evolution of \bafe. The dashed curves in panels (c) and (d) represent the prediction by the one-zone models (the same as those in IWP15).}
	\label{fig1}
	\end{center}
	\end{figure*}

In Figure \ref{fig1}, we present the chemical evolutions of five selected subhalos with different stellar masses for  case 1 (see Sec. 2.1). Their stellar masses are $10^4$, $10^5$, $10^6$, $10^7$, and $10^8 \, \eMsolar$ corresponding to those from the thinnest to thickest curves in different colors.

Figure \ref{fig1}a shows the age-metallicity relations of the five subhalos. The $\fe$ yields are those produced by CCSNe. Since CCSNe occur constantly and sufficiently even for low-mass subhalos (see IWP15), \feh\ \SW{in each subhalo} increases monotonically with time. For case 1, the star formation efficiency \ksf\ is a function of the subhalo mass, in which heavier subhalos have higher \ksf\ values. Therefore, a heavier subhalo shows a higher increasing rate of metallicity. Regarding case 2 (not shown here), \ksf\ is constant regardless of the subhalo mass and hence the age-metallicity relations are the same for all subhalo models. These results are consistent with IWP15.

Figure \ref{fig1}b shows the cumulative number of NSMs computed stochastically, which occur in each subhalo. In the subhalos with masses above $10^7 \, \eMsolar$, the first NSM occurs at $0.01$--$0.02 \, \gyr$. 
\SW{This is a consequence of the fact that} these subhalos contain a \SW{large} enough number of \SW{massive} stars for triggering even rare \SW{(short-lived)} NSMs. After $0.1 \, \gyr$, the cumulative \Nnsm\ quickly increases owing to the long-lived NSMs. 
For the heaviest subhalo, the cumulative \Nnsm\ increases smoothly unlike the others because the NSM occurrence rate is high. For the subhalos with masses below $10^6 \, \eMsolar$, the numbers of stars are too small to experience short-lived NSMs. The NSMs occurring at $0.1$--$0.3 \, \gyr$ are due to the long-lived NSMs. In any case, NSMs do not necessarily occur at $t = 0.1 \, \gyr$ because of the stochastic nature. The probabilistic effects appear more clearly in the subhalos with lower masses. The total number of \Nnsm that occur in these subhalos with masses of $10^5 \, \eMsolar$  and $10^4 \, \eMsolar$ are only 3 and 1, respectively.

Figure \ref{fig1}c shows the time evolutions of \bah\ in the five subhalos. We compare our results for these subhalos with the average evolutions that are calculated from the one-zone models (the same as those in IWP15). When the first NSM occurs in each subhalo, \bah\ becomes greater than the average value. \bah\ stays constant when no NSM occurs. The enhancement is stronger for a lower-mass subhalo\SW{, because} a lower mass of ISM leads to a greater ratio of $\ba$ to gas and hence a larger \bah. 

Figure \ref{fig1}d shows the time evolutions of \bafe, which are the combinations of those in Figures \ref{fig1}a and c. The amount of Ba is enhanced by NSMs. When no NSM occurs, the Ba abundance remains constant. However, CCSNe occur continuously,  and the amount of Fe increases. As a result, \bafe\ shows a monotonic decline until the next NSM event.
%
%


\subsection{$\erfe$ vs. $\efeh$ in Subhalos}
\label{subsec:case12}

\begin{figure*}
    \begin{center}
    \includegraphics[width=0.8 \textwidth]{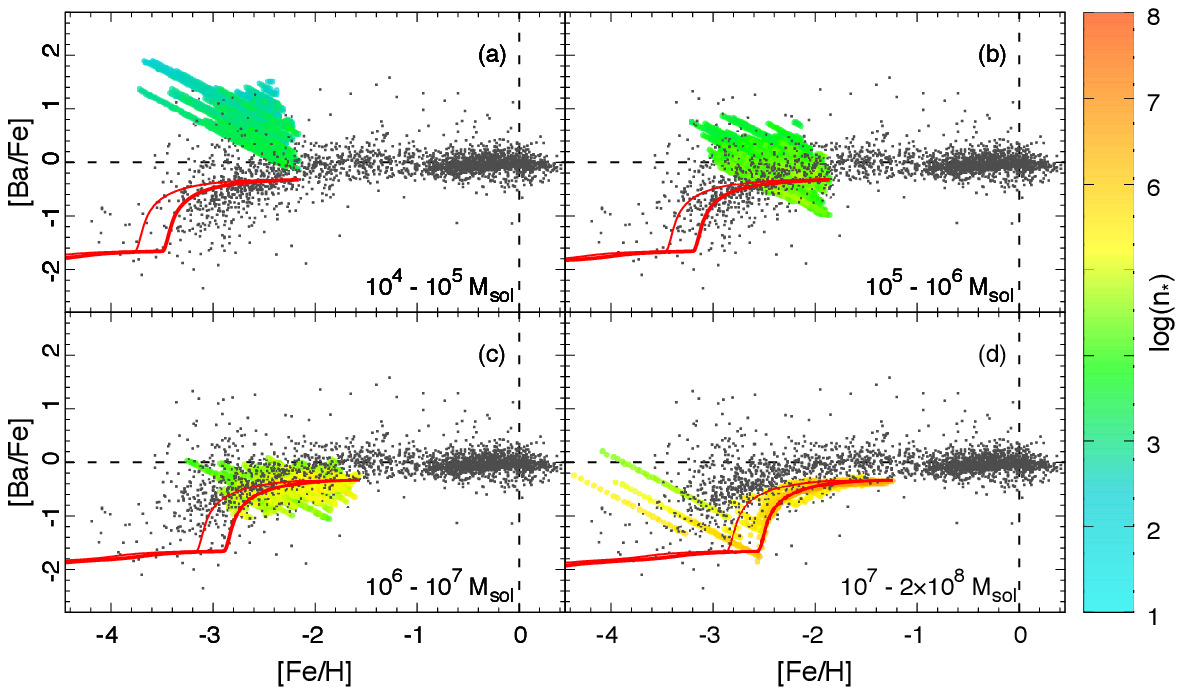}
	\caption{\bafe\ vs. \feh\ for case 1 in individual subhalos. The plotted subhalo mass ranges are (a) $10^{4}$--$10^{5} \,\eMsolar$, (b) $10^{5}$--$10^{6} \,\eMsolar$, (c) $10^{6}$--$10^{7} \,\eMsolar$, and (d) $10^{7}$--$2\times10^{8} \,\eMsolar$. The colored areas  represent the number distributions of stars in logarithmic scale. The solid curves in red show the average abundance ratios (the same as the models in IWP15), where the thinner (thicker) curve corresponds to the lowest (highest) mass subhalo. The gray dots show the abundance ratios in the observed Galactic halo stars taken from the SAGA database (\citealt{SAGA, Suda+11}; excluding the upper  limits and those of carbon-enhanced stars $(\ecfe \gtrsim 0.5)$). The dashed lines indicate the solar values.}
	\label{fig2}
     \end{center}
	\end{figure*}

Figure \ref{fig2} shows the chemical evolutions of $\ba$ in subhalos for case 1. The subhalos are grouped according to the mass ranges (a) $10^{4}$--$10^{5} \,\eMsolar$, (b) $10^{5}$--$10^{6} \,\eMsolar$, (c) $10^{6}$--$10^{7} \,\eMsolar$, and (d) $10^{7}$--$2\times10^{8} \,\eMsolar$. Each panel consists of several subhalos plotted together. A subhalo that experiences no NSM does not appear in the figure because there is no \rp\ enrichment. Therefore, Figure \ref{fig2}a shows only 48 out of 741 subhalos; Figure \ref{fig2}b shows 79 out of 147 subhalos; and Figures \ref{fig2}cd show all subhalos with the corresponding masses. In Table~3, we summarize the numbers of \rp-enriched subhalos that  appear in the figures.
%
%
	\begin{deluxetable}{ccccc}
	\tablecolumns{5}
	\tablewidth{0pc}
	\tablecaption{Number of \rp-enriched subhalos}
	\tablehead{
		\colhead{Panel} &
		\colhead{Mass Range} &
		\colhead{\Nsub}  &
		\colhead{} &
		\colhead{\hspace{-12mm} $\eNsub (\eNnsm \geq 1)$} \\
			\cline{4-5} \\ 
			\colhead{} &
			\colhead{(\Msolar)} &
			\colhead{} &
			\colhead{Case 1} &
			\colhead{Case 2}
		}
	\startdata
	(a) & $10^4$ -- $10^5$ & $741$ & $48$ & $72$ \\
	(b) & $10^5$ -- $10^6$ & $147$ & $79$ & $80$ \\
	(c) & $10^6$ -- $10^7$ & $29$ & $29$ & $29$ \\
	(d) & $10^7$ -- $2\times10^8$
			      		    & $7$ & $7$ & $7$ \\
	\enddata
	\label{tab:plottedsub}
	\end{deluxetable}

%
%

The thin and  thick (red) curves in each panel of Figure~\ref{fig2} represent the average values of abundance ratios for the subhalos with the lowest and highest masses, respectively (the same as the models in IWP15). In the beginning, $\ba$ is produced only by the rare short-lived NSMs and hence \bafe\ remains almost constant regardless of \feh. When the more numerous long-lived NSMs start contributing, \bafe\ begins to increase.
The distributions of \bafe\ calculated stochastically deviate from these average evolutionary curves. As already mentioned, not all subhalos experience NSMs, in particular for the low-mass groups (Figures \ref{fig2}ab). For example, only 48 out of 741 subhalos with the masses $10^4$--$10^5 \, \eMsolar$ experience NSMs. Thus, most of the subhalos $(\approx 94\%)$ do not appear in the figure. On the other hand, the rest of the subhalos appear, showing very high \bafe\ values. For massive subhalos ($\ge 10^6\, M_\odot$), most of the stars are concentrated near the average curves (Figure \ref{fig2}cd). Note that all of these massive subhalos have experienced NSMs.

    \begin{figure*}
    \begin{center}
    \includegraphics[width=0.8\textwidth]{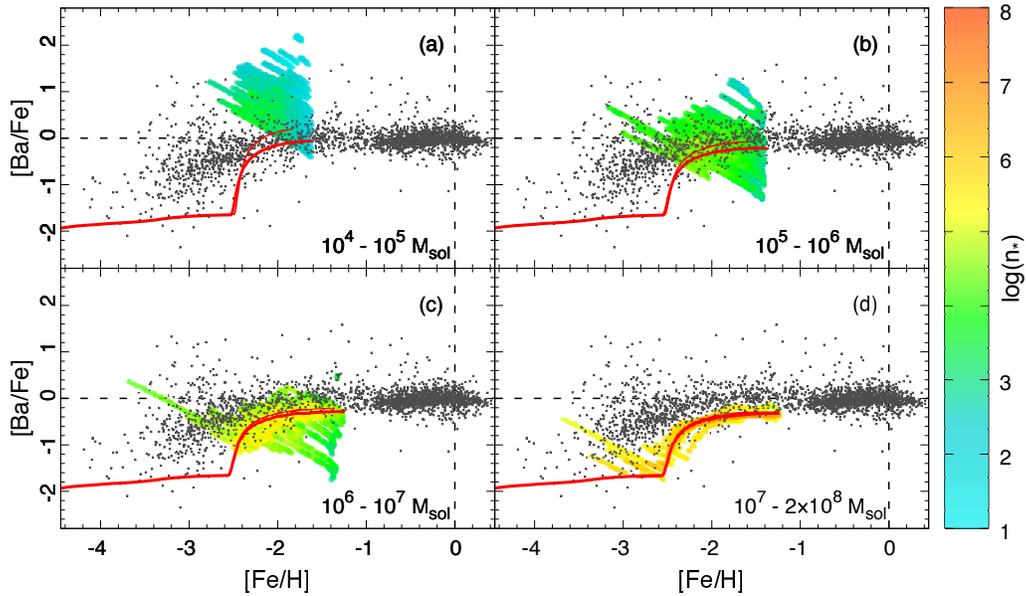}
	\caption{Same as Figure \ref{fig2}, but for case 2.}
	\label{fig3}
    \end{center}
    \end{figure*}

In general, once a single NSM occurs in a given subhalo, the evolution of \rfe\ begins with a high value that decreases monotonically with increasing metallicity until the next NSM event, as discussed in section \ref{subsec:chemev}.
As clearly seen in the average abundance ratios, the long-lived NSMs occur at lower metallicities for less massive subhalos. The stars in lower-mass subhalos have lower metallicities on average. Massive subhalos also have stars at low metallicities, but their \rpes\ ($\ba$) originate from the short-lived NSMs.


Figure \ref{fig3} shows the chemical evolutions of $\ba$ for case 2. The enhancements of \bafe\ in the average values start at $\efeh \sim -2.5$ for all subhalos, regardless of their masses. The stochastically calculated models also show \bafe\ enhancements from similar metallicities. Enhancements are also visible at lower metallicities, but the number of such subhalos is small, which are polluted by the rare short-lived NSMs. At $\efeh \sim -1.5$, the \bafe\ values are rather high for low-mass subhalos (Figures \ref{fig3}ab). This is due to the small ISM masses at late times of their evolutions by strong gas outflow (section~2.1).

\subsection{Clustering of Subhalos}

According to the hierarchical structure formation scenario, our Galactic halo has been formed from the clustering of subhalos \citep[e.g.,][]{Prantzos08}. In this context, one can examine the chemical evolution of the Galactic halo as the ensemble of the subhalos discussed in section~3.2. 

	\begin{figure}
	\epsscale{1.2} 
	\plotone{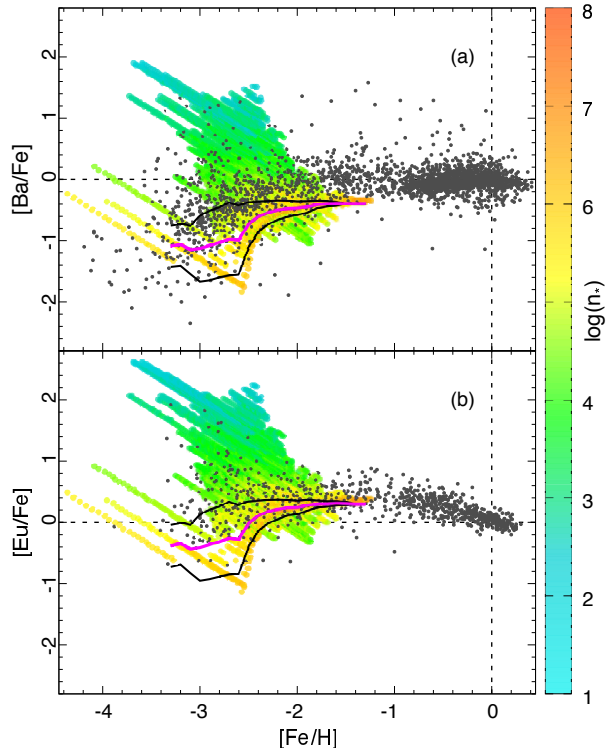}
	\caption{Chemical evolution in the Galactic halo as the ensemble of all subhalos for case~1, where (a) [Ba/Fe] and (b) [Eu/Fe] are displayed as functions of [Fe/H]. The colored areas  represent the number distributions of stars in logarithmic scale. The thick (magenta) and thin (black) solid curves indicate the mean values and the standard deviations of the \rfe\ distributions in each $0.1 \, \dex$ bin of \feh, respectively. The dots show the abundance ratios in the observed Galactic halo stars taken from the SAGA database (\citealt{SAGA, Suda+11}; excluding  upper limits and those of carbon-enhanced stars $(\ecfe \gtrsim 0.5)$). The dashed lines indicate the solar values.}
	\label{fig4}
	\end{figure}



Figure \ref{fig4} shows the chemical evolutions of $\ba$ and $\eu$ relative to $\fe$ in the Galactic halo as functions of \feh, as the ensemble of all subhalos for case 1. Both [Ba/Fe] and [Eu/Fe] show large star-to-star dispersions of $\sim 3 \, \dex$ at $\efeh \lesssim -2.5$, which converge with increasing metallicity. The mean value (taken in each 0.1~dex bin of [Fe/H]; magenta) of the abundance ratios increases with metallicity. The standard deviation (black) from the mean value is maximal at $\efeh \sim -3 \, \dex$, in which the spread appears consistent with the observed dispersion of stars. All of these aspects are in good agreement with the observed trends of the Galactic halo stars ($\feh\ < -1$). Note that the mean value of \bafe\ becomes smaller than that of the stellar abundance ratios at $\efeh \gtrsim -2.5$ because we do not consider the  contributions from the $s$-process \citep[e.g.,][]{Prantzos+18}.

By comparing Figures \ref{fig2} and \ref{fig4}a, it is clear that the highly \rp-enhanced stars in the Galactic halo originate from the low-mass subhalos $(\eMsub \lesssim 10^6 \, \eMsolar)$. On the other hand, the Galactic halo stars with subsolar \rfe\ values mostly come from the heavier subhalos with masses greater than $10^6 \, \eMsolar$. The presence of such a large dispersion in \rfe\ at $\feh\ \sim -3$ cannot be reproduced by the evolutions of averaged abundance ratios (red lines in Figure~2; the same as the models in IWP15), which is obviously due to a stochastic nature of NSM events in subhalos. However, the large standard deviation between $\feh\ \sim -3$ and $-2.5$ (Figure~4) is mainly due to the mass-dependent star formation efficiency for case~1, $\eksf \propto {\eMsub}^{+0.3}$, not to the stochastic occurrences of NSMs. As shown in IWP15, the subhalos evolve at different rates, reaching different metallicities at a given time (or the same metallicity at different times, thus with different NSM rates). As a consequence, the stars of a given [Ba/Fe] are distributed in different metallicity regions when all subhalos are displayed (see the red lines in Figure~2).
%
%

\begin{figure}
	\epsscale{1.2} 
	\plotone{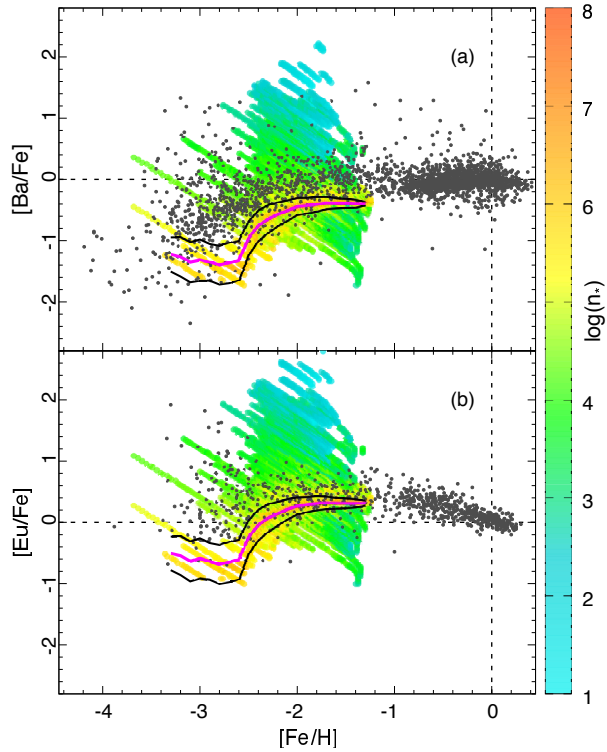}
	\caption{Same as Figure \ref{fig4}, but for case 2.}
	\label{fig5}
	\end{figure}

Figure \ref{fig5} shows the results for case 2. 
\SW{The calculated stellar \rfe\ values with large star-to-star scatters appear at the metallicity $\efeh \gtrsim -2.5$, which is substantially higher than $\efeh \sim -3$ in the observations.} 
The mean value of our result shows a sudden increase at $\feh\ \sim -2.5$; however, the observed \rfe\ values gradually increase from $\efeh \sim -3$ to $-2$. Unlike case 1, the standard deviation overall is small despite its large star-to-star scatter. The reason is the constant star formation efficiency for case~2, $\eksf = 0.20 \, {\gyr}^{-1}$, which keeps all subhalos on the same evolutionary path of \rfe\ (see the red lines in Figure~3), resulting in a small standard deviation.
%
%



We find that our model of case 1 is successful in explaining the observational aspects of the \rp\ abundances in Galactic halo stars: the large \rfe\ dispersion with highly \rp-enhanced stars at $\efeh \lesssim -2.5$ and the mean value of \rfe, which gradually increases with metallicity from $\efeh \sim -3$. Therefore, the model of subhalos with $\eksf \propto {\eMsub}^{+0.3}$ appears suitable to describe the chemical evolution of the Galactic halo, and thus we focus on case 1 in the subsequent discussion.


\subsection{NSM Coalescence Timescales}

In the previous subsections, the NSM coalescence timescale \tnsm\ was assumed as $100 \, \myr$ for $95\%$ of the NSMs and $1 \, \myr$ for the rest. In this subsection, we test other choices of \tnsm.

\begin{figure}
	\epsscale{1.2} 
	\plotone{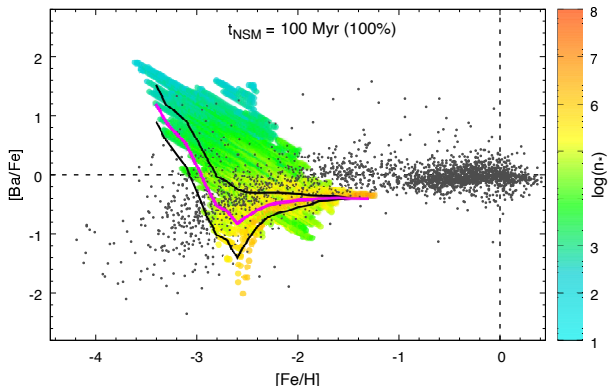}
	\caption{Same as Figure~4a, but for the coalescence timescale $\etnsm = 100 \, \myr$ for all NSMs.}
	\label{fig6}
	\end{figure}

In order to examine the case of little contribution from the short-lived NSMs, we first calculate a model using only the long \tnsm, that is, $\etnsm = 100 \, \myr$ for all NSMs. Figure \ref{fig6} shows the calculated \bafe\ evolution in the Galactic halo for case 1. The Galactic halo has the metal-poor stars with subsolar \bafe\ values at metallicities less than $-3$, which cannot be explained solely with $\etnsm = 100 \, \myr$ in our model.
%
%

\begin{figure}
	\epsscale{1.2} 
	\plotone{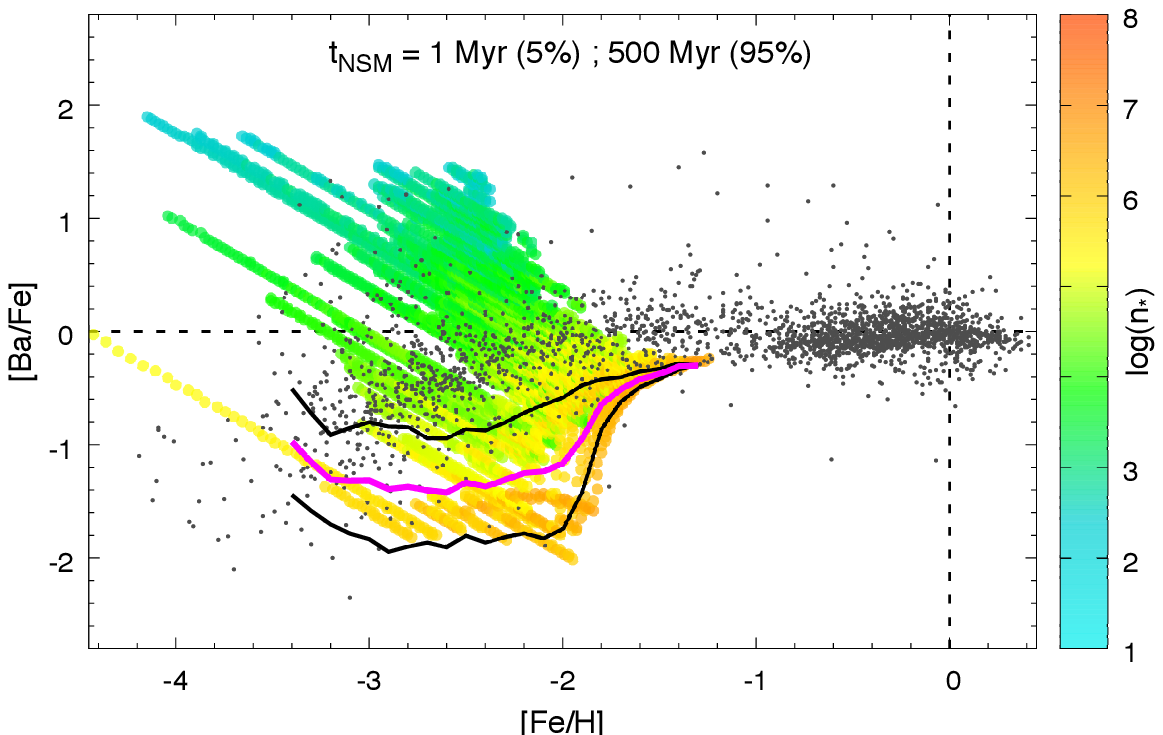}
	\caption{Same as Figure~4a, but for the coalescence timescales $\etnsm = 1\,\myr$ and 500~Myr with the corresponding fractions of $5\%$ and $95\%$, respectively.}
	\label{fig7}
	\end{figure}

Since population synthesis studies as well as the observation of binary neutron stars infer the NSM coalescence timescales $\ge 100 \, \myr$, we next test the case of a longer \tnsm. Figure \ref{fig7} shows the calculated \bafe\ evolution in the Galactic halo, where the coalescence timescale of the long-lived NSM is set to $500 \, \myr$ for $95\%$ of the NSMs and the rest with $\etnsm = 1 \, \myr$. The increase of \bafe\ due to the long-lived NSMs starts at higher metallicity $(\efeh \sim -2)$ and overproduces stars with $\ebafe < -1$. Also, the star-to-star scatter becomes largest at $\efeh \sim -2.5$, whereas observations show its large dispersion at $\efeh \sim -3$. This implies that a coalescence timescale appreciably greater than 100~Myr cannot represent the majority of NSMs.
%
%


\subsection{$\emgfe$ vs. $\efeh$}

\begin{figure}
	\epsscale{1.2} 
	\plotone{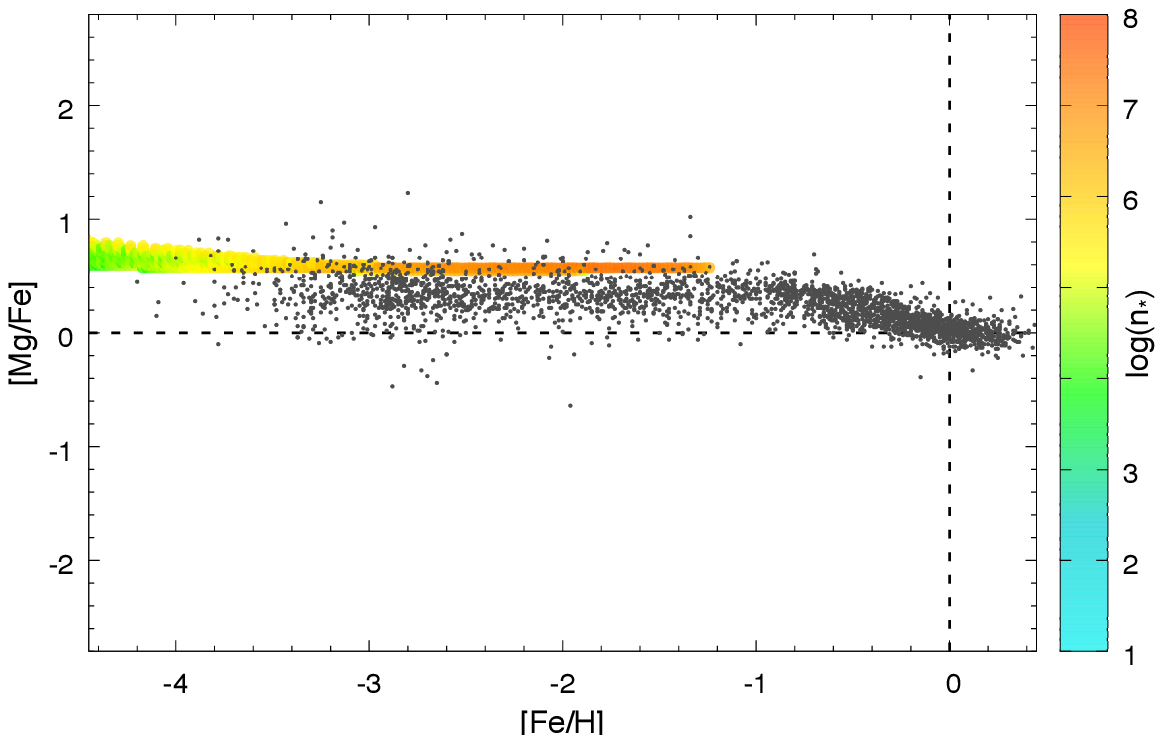}
	\caption{Same as Figure~4, but for  \mgfe\ vs. \feh.}
	\label{fig8}
	\end{figure}

The abundance ratios of \rpes\ in the metal-poor halo stars show large dispersions. In contrast, the abundance ratios of $\alpha$-elements such as Mg show a small spread of $\sim 1 \, \dex$ (see the gray dots in Figure \ref{fig8}). \SW{Moreover, a well-selected homogeneous sample shows little intrinsic scatter in [Mg/Fe] at low metallicity \citep[within a total range of 0.2~dex with a standard deviation of 0.06~dex for 23 stars;][]{Arnone+05}.} \mgfe\ appears uniform throughout the whole metallicity range until type Ia supernovae start to increase the $\fe$ abundance after $\efeh \sim -1$. \SW{This contradicts the inhomogeneous chemical evolution models with inefficient ISM mixing, in which the variation of Mg/Fe in CCSNe yields remains at low metallicity \citep[see, e.g.,][]{Argast+02}. Recent hydrodynamical studies show that the implementation of substantial ISM mixing is fundamental to reproduce the observed little scatter in [Mg/Fe] \citep[e.g.,][]{Shen+15, Hirai+17b}. Each of the one-zone homogeneous subhalos in our model can be regarded as a limiting case of such efficient ISM mixing.} In this subsection, we test whether our model is also consistent with the observed behavior of $\alpha$-elements such as $\mg$.

The calculated $\mg$ abundance evolution relative to $\fe$ as a function of \feh\ for all subhalos is shown in Figure \ref{fig8}. Results show very small star-to-star scatters of \mgfe\ for $\efeh < -1$, being in agreement with \SW{the observational trend. The higher [Mg/Fe] values than the mean of the measured values ($\approx 0.5$) reflect the CCSN yields adopted in this study}. Every subhalo is assumed to have a well-mixed ISM, meaning that the elements made by CCSNe from all the progenitor mass range are mixed uniformly. Therefore, every subhalo shows similar \mgfe\ values over a wide range of metallicities. The resulting abundance ratios \SW{in the Galactic halo as an ensemble of these subhalos} display little dispersion.
%
%



\section{Chemical evolution of S\lowercase{r}}
\label{sec:strontium}


Sr is one of the \SW{light trans-iron} elements\footnote{Light trans-iron elements from Zn ($Z = 30$) to Zr ($Z = 40$), including Sr, are not necessarily made by neutron capture but in nuclear equilibrium in the neutron-rich ejecta of CCSNe \citep{Wanajo+11, Wanajo+18}.}, showing a large dispersion of \srfe\ in the metal-poor stars of the Galactic halo (see observational data in Figure \ref{fig10}) similar to Ba and Eu. However, the light-to-heavy abundance ratios  \srba\ also show a dispersion with a large number of stars having values greater than the solar ratio (see the observational data in Figure \ref{fig11}). Such a trend cannot be explained from a single \rp\ site that produces both Sr and Ba with the solar \rp\ ratio. This fact implies the presence of a ``weak" \rp\ \citep{Wanajo+06} that makes Sr but little Ba, in addition to the ``main" \rp. 
HD~122563 $(\efeh = -2.7)$ is such a star with the signature of a possible weak \rp\ \citep{Honda+06}, exhibiting a descending trend of abundances toward heavier neutron-capture elements (see also \citealt{Aokim+17}).


In this section, we examine our model (case 1) for the evolution of $\sr$ by taking into account the contributions from both the main and weak $r$-processes. The assumed site of the main \rp\ is NSMs as in the previous sections. Note that such a modification of our model does not affect the evolutions of Ba or Eu in section~3. For the weak \rp\, we assume the stars with $10$ -- $11 \, \eMsolar$, at the low-mass end of CCSN range, which are suggested as  the primary sources of light trans-iron elements including Sr \citep{Wanajo+11, Wanajo+18}. The solar \rp\ abundance of $\sr$ by mass is $16$ times greater than that of $\eu$. Thus, we define $y_{\sr,m}$ and $y_{\sr,w}$, the yields of $\sr$ from the main and weak \rp es, respectively, which satisfy
	\begin{equation}
	16 \, \eNnsm \, y_{\eu} = \eNnsm \, y_{\sr,m} + N_w \, y_{\sr,w} ,
	\label{eq:ysr1}
	\end{equation}
\SW{where $N_w$ is the number of CCSNe from the range $10$ -- $11 \, \eMsolar$.} Observations show that the highly \rp-enhanced stars such as CS~22892-052 and CS~31082-001 have somewhat smaller light-to-heavy \rp\ abundance ratios than that of the solar \rp\ abundances (\citealt{Sneden+08, Siqueira+14}; see also \citealt{Ji+18} for the stars in Ret~II). For simplicity, we assume that half of the $\sr$ abundances come from the main \rp\ and the other half from the weak \rp, i.e.,
	\begin{equation}
	\eNnsm \, y_{\sr,m} = N_w \, y_{\sr,w} .
	\label{eq:ysr2}
	\end{equation}
The yields of $\sr$ from the main and weak \rp es are thus derived from Equations~(\ref{eq:ysr1}) and (\ref{eq:ysr2}). The parameters related to $\sr$ yields are summarized in Table~4.
%
%
	\begin{deluxetable}{rcr}
	\tablecolumns{3}
	\tablewidth{0pc}
	\tablecaption{Strontium Yields}
	\tablehead{
		\colhead{Site} &
		\colhead{Progenitor Mass Range} &
		\colhead{Yield} \\
			\colhead{} &
			\colhead{(\Msolar)} &
			\colhead{(\Msolar)}
		}
	\startdata
	CCSN (weak $r$) & $10$ -- $11$ & $2.0 \times 10^{-6}$ \\
	NSM (main $r$) & $10$ -- $60$ & $3.2 \times 10^{-4}$ \\
	\enddata
	\label{tab:sr}
	\end{deluxetable}

\begin{figure*}
    \begin{center}
	\includegraphics[width=0.8\textwidth]{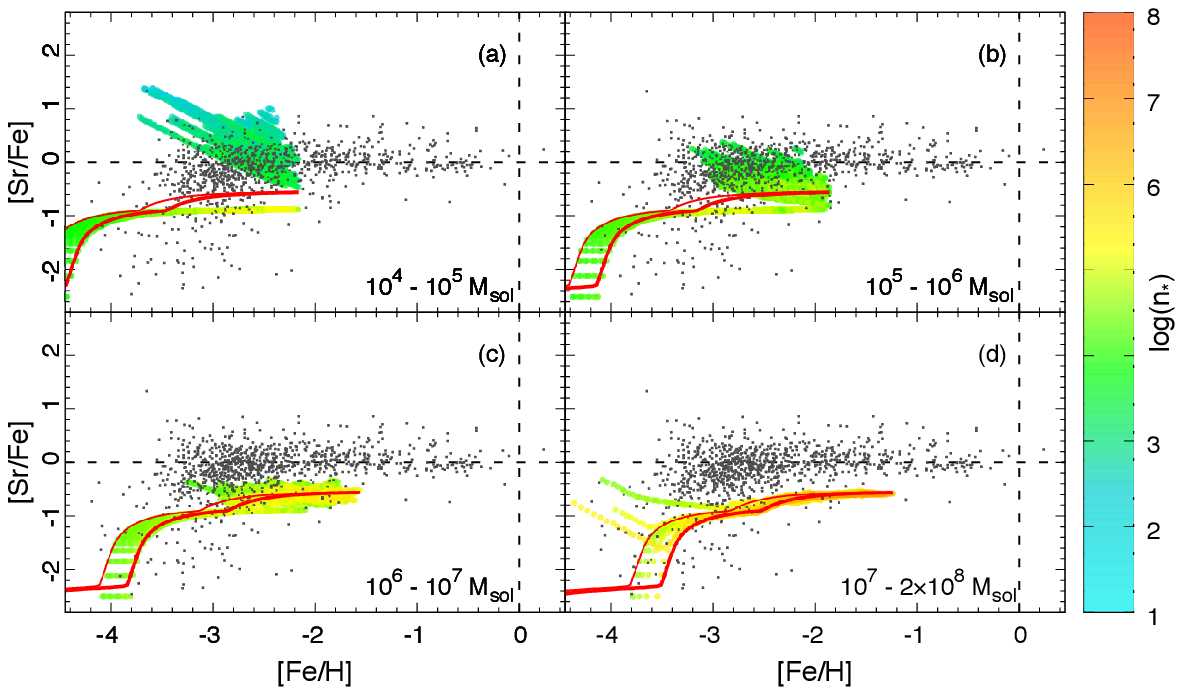}
	\caption{Same as Figure \ref{fig2}, but for [Sr/Fe] vs. [Fe/H].}
	\label{fig9}
    \end{center}
	\end{figure*}
\begin{figure}
	\epsscale{1.2} 
	\plotone{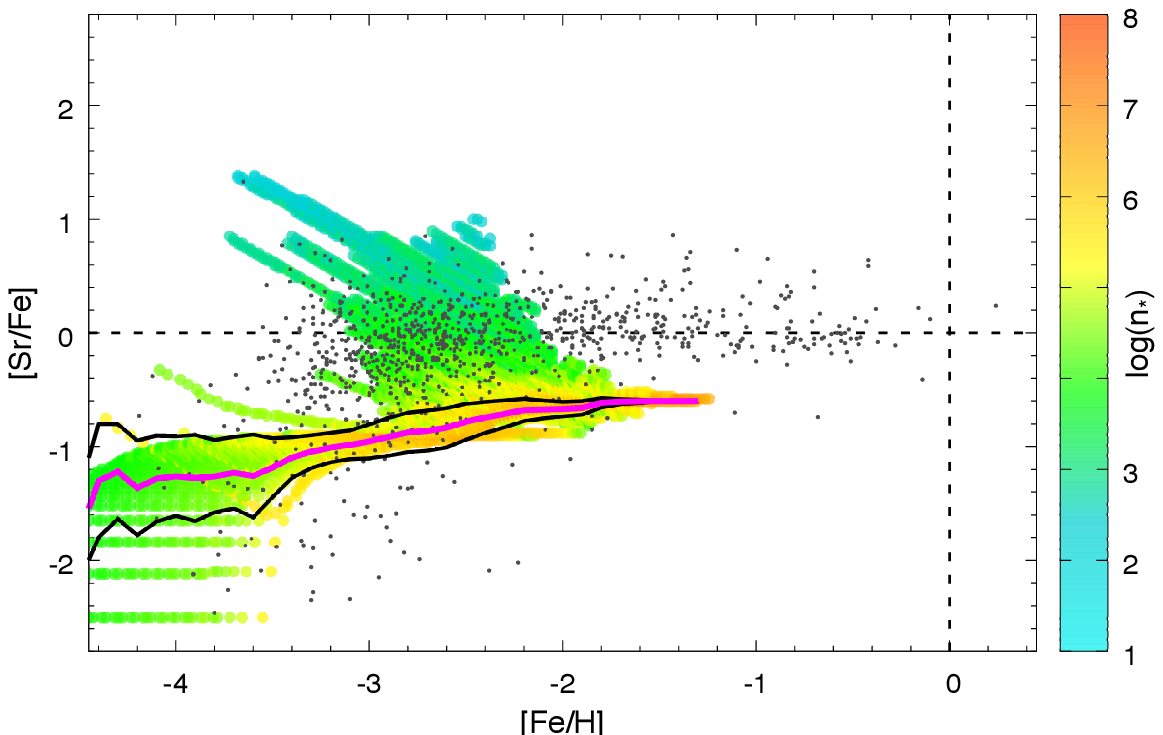}
	\caption{Same as Figure~4, but for \srfe\ vs. \feh.}
	\label{fig10}
	\end{figure}
\begin{figure}
	\epsscale{1.2} 
	\plotone{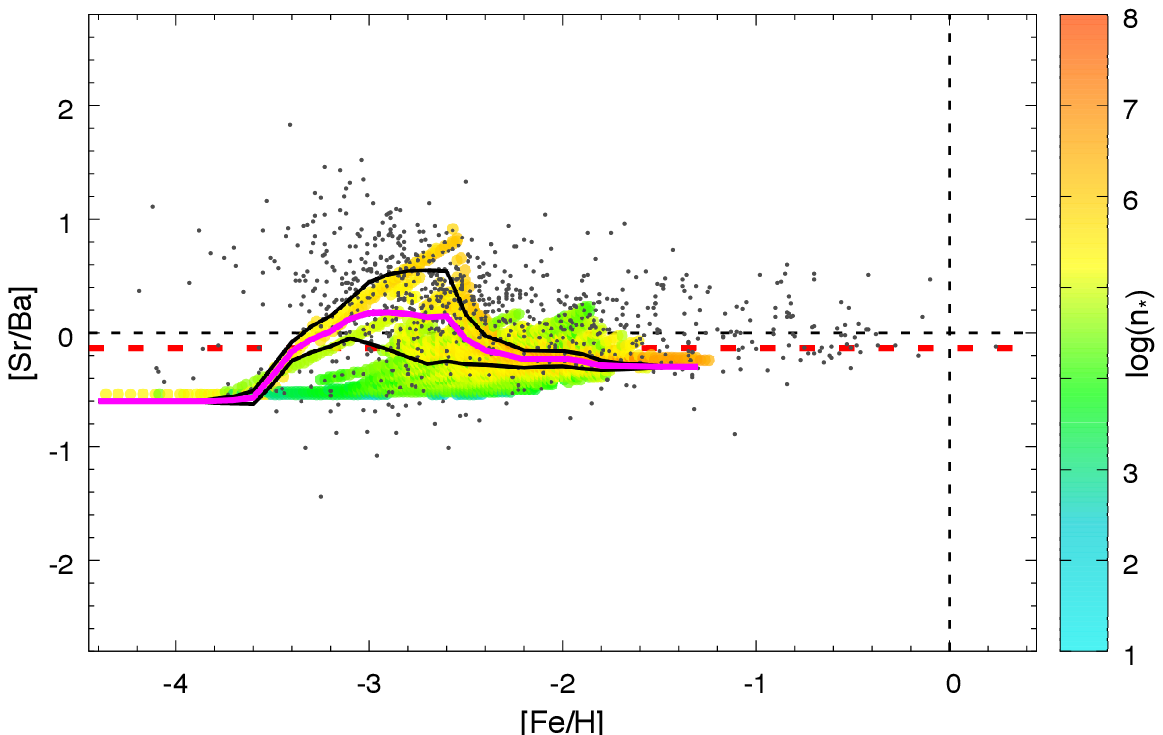}
	\caption{Same as Figure~4, but for \srba\ vs. \feh. The thick dashed horizontal line (red) indicates the solar ratio of \srba\ for the \rp\ component ($11\%$ for $\ba$ and $15\%$ for $\sr$; \citealt{Burris+00,Sneden+08}).}
	\label{fig11}
	\end{figure}

We present the chemical evolutions of $\sr$ relative to $\fe$ as functions of \feh\ (case~1) for different mass subhalos in Figure \ref{fig9}. The average enhancement of \srfe\ occurs first by the short-lived NSMs up to $\esrfe \sim -2.4$ \SW{along with an increase of metallicity to [Fe/H] $\sim -3$}. Afterward, the weak \rp\ increases the $\sr$ abundance ratio up to $\srfe \sim -1$, followed by the enrichment due to the long-lived NSMs.
The weak \rp\ (low-mass CCSNe) is responsible for the enrichment of the stars with $\esrfe \lesssim -1$, whereas the higher \srfe\ stars are entirely due to the main \rp\ (NSMs). As discussed in sections \ref{subsec:chemev} and \ref{subsec:case12}, NSMs highly enhance the \srfe\ values, especially in a fraction of low-mass subhalos. In contrast to NSMs, \SW{low-mass} CCSNe occur uniformly in all subhalos, making no star-to-star scatters or large enhancements of \srfe.

%
%

The chemical evolution of Sr as an ensemble of all subhalos is shown in Figure \ref{fig10}, which is compared with the abundances of Galactic halo stars. The observed trend of \srfe\ such as the large star-to-star scatter at $\efeh \sim -3$ with highly $\sr$ enhanced stars at $\esrfe \sim 1$ is well reproduced as those of Ba and Eu discussed in section~3. The mean calculated value of [Sr/Fe] at $\feh\ \gtrsim -2.5$ is lower than the solar value because we exclude the contributions from the $s$-process.
%
%


We also present the evolution of $\esrba$ in Figure \ref{fig11}. As found in the evolution of \srfe, the average enhancement of \srba\ to a constant value of $-0.6$ occurs first by the short-lived NSMs. The rise of \srba\ at $\efeh$ \SW{$\lesssim -3$} with a large dispersion is due to the production of $\sr$ (without Ba) by the weak \rp\ in low-mass CCSNe. Afterwards, the long-lived NSMs start producing both Sr and $\ba$, which leads to a convergence of [Sr/Ba] values \SW{at [Fe/H] $\gtrsim -3$}. Note that the \srba\ value is constant at $-0.54$ with our adopted yields for NSMs, meaning the higher values are purely due to the weak \rp. We also find an overall agreement of our model with the observed trend of [Sr/Ba], that is, the large star-to-star scatter of the abundances at $\feh\ \sim -3$ with few stars below $\srba\ \sim -0.6$. It is emphasized that the reasonable agreement here is due to the additional sources of Sr, low-mass CCSNe in our model, in addition to NSMs.
%
%


\section{Chemical evolution of Ultra-Faint Dwarf Galaxies}
\label{sec:ufd}


Ultra-faint dwarfs (UFDs) are small galaxies in mass and size with low luminosities. Their faintness makes their discovery as well as their spectroscopic study difficult. Up to date, about 10 UFDs have been discovered around our Galaxy. They are metal-poor (e.g., \citealt{Kirby+13}), mostly showing very low \rfe\ (e.g., \citealt{Francois+07, Koch+13, Frebel+14, Francois+16}). However, one of the UFDs, Reticulum II (Ret II), has highly \rp-enhanced stars \citep{Ji+16_Nature, Roederer+16, Ji+16_Sr, Ji+18}. The fact that Ret II is one such peculiar galaxy out of about 10 UFDs is reminiscent of our result for low-mass subhalos discussed in section~3. This similarity motivates us to explore the chemical evolutions of UFDs by adopting our models of low-mass subhalos. In this subsection, we apply our subhalo model to the chemical evolution of Ret~II. 

As discussed in section \ref{subsec:chemev} and \ref{subsec:case12}, the relative abundance ratio in a low-mass subhalo sizably deviates from the average evolutionary curve. This is due to the small number of NSMs occurring in each subhalo. As a result, the value of maximum \Nnsm\ divided by the mean \Nnsm\ for a given mass range (see Table~1) is greater for lower-mass subhalos.

The stellar mass in a typical UFD galaxy is $\sim 10^3$ -- $10^4 \, \eMsolar$. Provided that UFDs evolve the same as subhalos, we consider the lowest-mass range in our model (case~1), $10^{4.0}$ -- $10^{4.1} \, \eMsolar$, in which only 5 out of $138$ subhalos experience (only single) NSMs (Table~5).
%
%
	\begin{deluxetable}{cccc}
	\tablecolumns{5}
	\tablewidth{0pc}
	\tablecaption{Properties of $r$-process-enriched UFD Models}
	\tablehead{
		\colhead{Model} &
		\colhead{\feh\ at} &
		\colhead{\Nnsm} &
        \colhead{fraction of} \\
            \colhead{} &
			\colhead{Enhancement} &
            \colhead{} &
            \colhead{Enriched Stars}
		}
	\startdata
    SH1 & $-3.59$ & $1$ & $0.846$ \\
    SH2 & $-3.35$ & $1$ & $0.750$ \\
    SH3 & $-2.68$ & $1$ & $0.225$ \\
    SH4 & $-2.53$ & $1$ & $0.0874$ \\
    SH5 & $-2.45$ & $1$ & $0.0187$ \\
	\enddata
	\label{tab:ufd}
	\end{deluxetable}

\begin{figure*}
	\includegraphics[width=1.0 \textwidth]{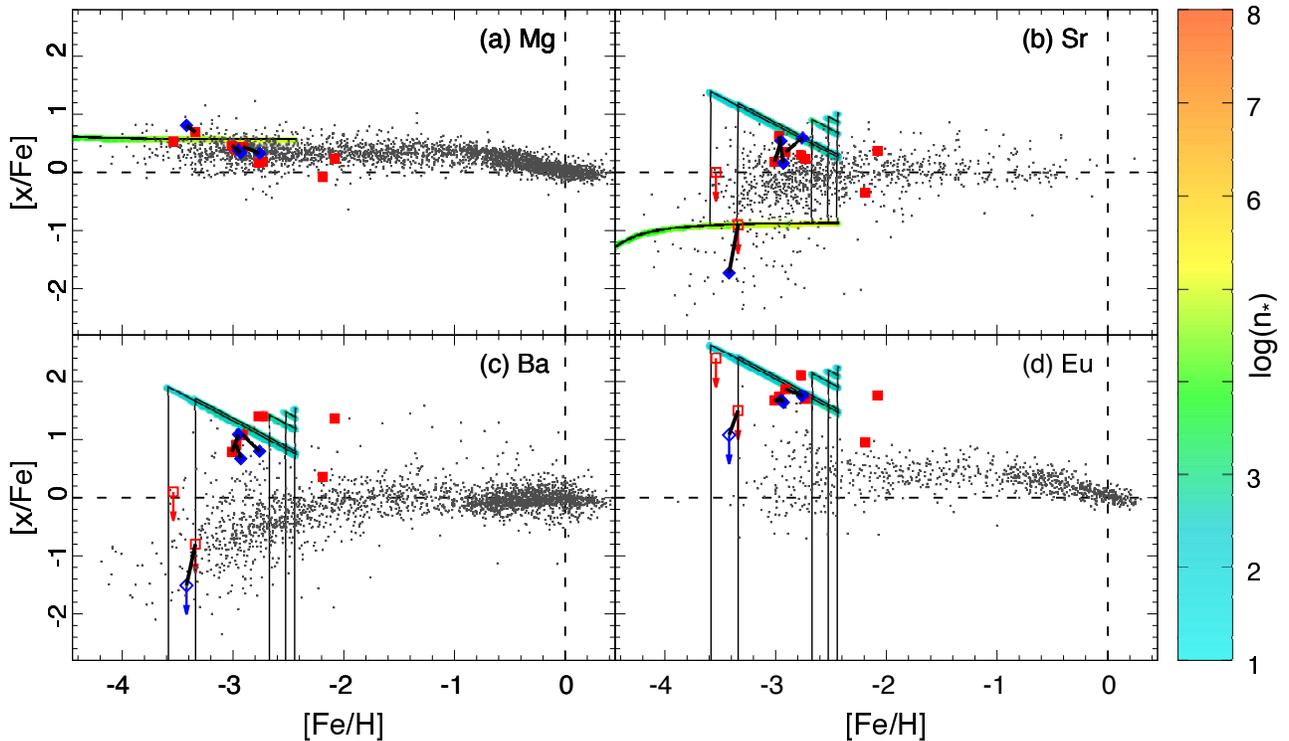}
	\caption{Chemical evolutions of [$x$/Fe] as functions of \feh\ in our 138 UFD models (subhalos in the range of $M_\mathrm{sub}/M_\odot = 10^{4.0}$--$10^{4.1}$), where $x$ is the element (a) Mg, (b) $\sr$, (c) $\ba$, and (d) $\eu$.  The colored areas represent the numbers of stars in logarithmic scale. The evolutionary track of a single galaxy is represented by a thin line (overlapped for all galaxies in (a)). Five UFD models (see the first column in Table~5) show the enrichment of \rp\ elements in panels (b), (c), and (d), while the other 133 models do not appear in panels (c) and (d). The filled squares (red) and diamonds (blue) show the data of the Ret II stars from \citet{Ji+16_Sr} and \citet{Roederer+16}, respectively. The open symbols are their upper limits. The same stars are connected by the thick lines. Model~SH2, which starts the enrichment of \rp\ elements at $\feh\ = -3.53$, well reproduces the evolutions of all elements presented here. The small dots show the observed abundances in the Galactic halo stars taken from the SAGA database (\citealt{SAGA, Suda+11}; excluding upper limits and those of carbon-enhanced stars $(\ecfe \gtrsim 0.5 \, \dex)$). The dashed lines indicate the solar values.}
	\label{fig12}
	\end{figure*}

In Figure \ref{fig12} we compare the chemical evolutions of these subhalos with the spectroscopic data of Ret II stars. Each band made by successive colored circles represents the evolution of a single galaxy, which is guided by a thin line. For Mg (Figure~12a), the evolutions of all 138 galaxies are overlapped because of their common origins for Mg and Fe, i.e., CCSNe, and cannot be distinguished. For Sr (Figure~12b), the early evolutions at $\feh\ < -3.6$ are due to the weak \rp\ from low-mass CCSNe and thus are overlapped as well. However, at $\feh\ > -3.6$, five UFD models (see the first column in Table~5) exhibit the jumps of \rfe\ values at various metallicities owing to single NSM events. For Ba and Eu (Figures~12cd), only five UFD models appear as a result of NSMs.

We find that model SH2 that starts the enrichment of \rp\ elements at $\feh\ = -3.35$ \SW{qualitatively captures the evolutionary trend} of Ret~II, namely, the small dispersion of [Mg/Fe] as well as the enrichment of Sr, Ba, and Eu for the seven stars at $\feh\ \gtrsim -3$ (and the low-level abundances or upper limits for the other two stars at $\feh\ \lesssim -3$). In model~SH~2 the \rp-deficient and \rp-enhanced stars are formed before and after the (single) NSM event, respectively. The fraction of \rp-enhanced stars in model~SH2 is 0.750 (the fourth column in Table~5), which is in good agreement with that in Ret~II ($7/9 = 0.778$). \SW{However, the models considered here reach the end of evolution at [Fe/H] $= -2.4$ and thus do not account for the presence of two stars at [Fe/H] $=-2.2$ and $-2.1$.}

Given that the evolutions of UFD galaxies are represented by our models of the lowest-mass subhalos, the fact that 5 out of $138$ subhalos experience NSMs implies a chance of several percent to discover Ret II-like galaxies with respect to all observed UFDs. Our result also suggests that UFDs with various fractions of \rp-enhanced stars will be discovered in future observations, e.g., 2 out of 10 stars with the enrichment of \rp\ elements (model~SH3; see the fourth column in Table~5).
%
%


\SW{One should keep in mind, however, that a trial here is for a qualitative purposes only. The least massive subhalo in our model has $10^4\, M_\odot$, which corresponds to intermediate-mass UFDs. It is currently unclear if the scaling of the mass-metallicity relation extends to $\sim 10^3\, M_\odot$ for the majority of UFDs \citep{Kirby+13}. It also is unknown if the chemical evolution of such a low-mass system, probably with only a few episodes of star formation, can be approximated with our simple picture of a homogeneous and continuous chemical evolution.}


\section{Summary and Conclusion}
\label{sec:summary}

We have revisited the study of Galactic chemical evolution by IWP15 in order to investigate the role of NSMs as the dominant sources of \rp\ elements in the Galaxy. Our chemical evolution model was constructed on the basis of the hierarchical structure formation scenario in IWP15, in which the different mass ($M_\mathrm{sub}$) subhalos that formed the Galactic halo  had different star formation histories. The number of NSMs occurring in each subhalo was obtained with the Monte Carlo method. The star formation histories of the subhalos were determined from the observed mass-metallicity relation of the local dwarf galaxies, assuming the same correlation for both systems. In the framework of our simple model of galactic chemical evolution, this mass-metallicity relation leads to $\ofr/\sfr \propto {\eMsub}^{-0.3}$, where $\ofr$ and \sfr\ are the outflow rate and the star formation rate, respectively. We examined two extreme cases as in IWP15, in which either of the coefficients for SFR and OFR (Equations~(1) and (2)) were kept constant such that $\ekof = 1.0 \, \gyr^{-1}$ and $\eksf \propto {\eMsub}^{+0.3}$ for case 1 or $\eksf = 0.20 \, \gyr^{-1}$ and $\ekof \propto {\eMsub}^{-0.3}$ for case 2.

Our result shows that the observed properties of the enrichment histories of \rfe\ in the Galactic halo can be explained by assuming that NSMs are the sources of \rp\ elements. The adopted low SFR in low-mass subhalos for case~1 makes the occurrence of NSMs at low metallicities ($\feh\ \sim -3$) possible as shown by IWP15. In addition, the presence of the \rp-enhanced metal-poor stars in the Galactic halo is accounted for as a result of the large enhancement of \rfe\ by a single or a few NSMs in a small fraction of low-mass subhalos. However, our case 2 adopting a constant star formation efficiency ($k_\mathrm{SF}$) results in the enrichment of \rp\ elements at  higher metallicities, $\feh\ > -2.5$, as found in previous studies. We conclude, therefore, that the reality is closer to our case~1, i.e., star formation is less efficient in lower-mass subhalos while gas outflow only weakly depends on the subhalo masses, provided that NSMs are the dominant contributors of \rp\ elements in the Galaxy. 

The observed trend of \rfe\ in the Galactic halo can be mostly reproduced solely by long-lived NSMs with a coalescence timescale of $100 \, \myr$. However, a small fraction of short-lived (1~Myr in our model) NSMs appear to be necessary, which are responsible for explaining the presence of stars with subsolar \rfe\ values at $\feh\ \lesssim -3$. A test shows that a long coalescence timescale appreciably greater than 100~Myr, such as $500 \, \myr$, has difficulty in reproducing the enhancement of \rpes\ at low metallicities. This implies either that in reality the distribution of coalescence timescales (for the long-lived NSMs) has a sharp peak at $\sim 100$~Myr or the NSMs with longer timescales escape from subhalos because of neutron star kicks and do not contribute to Galactic chemical evolution \citep{Beniamini+16b, Safarzadeh+17}. In the future, localizations of NSMs in galaxies by identification of electromagnetic counterparts (kilonovae) of gravitational waves will provide us with information on the distribution of coalescence timescales.


It is important to note that our model naturally reproduces a large dispersion of abundance ratios (relative to Fe) for \rpes, but a small one   for intermediate-mass elements such as Mg; the latter are produced by the same sources and on the same timescales as Fe, while the former result from different sources, operating on very different timescales. This feature, which is in good agreement with spectroscopic results, is another important aspect of NSMs as sources of \rpes.

Our model is also successful in explaining the spectroscopic abundances of light neutron-capture elements such as $\sr$ when assuming an additional contribution (a weak \rp) from low-mass CCSNe. The resulting evolution of \srba\ as a function of metallicity reasonably reproduces the observed star-to-star scatters with higher [Sr/Ba] values than those predicted solely by the enrichment from NSMs. This supports the idea that there are extra sources (weak \rp) of light neutron-capture elements in addition to the (main) \rp.

Finally, our models of the least massive subhalos with $M_\mathrm{sub} \sim 10^4\, M_\odot$ well account for the observed nature of UFDs, namely, only 1 (Ret~II) out of about 10 such galaxies shows enhancement of \rpes. Moreover, the fact that seven out of observed nine stars in Ret II show enrichment of \rpes\ can be reasonably reproduced by such a lowest-mass subhalo model. This supports the idea that the UFDs are the leftovers of the building blocks that made the Galaxy. Our models also predict the presence of Ret~II-like UFDs with various fractions of \rp-enhanced stars. This will be tested by future spectroscopic explorations of UFD galaxies. \SW{However, it is currently unclear if the mass-metallicity relation can be applied to the bulk of UFDs. It also is cautioned that we applied our homogeneous and continuous chemical evolution model to such a small system that probably had only a few episodes of star formation. Obviously, further studies of UFDs will be necessary from both observational and theoretical sides.}

\SW{It should be noted that our study neglected a spatial inhomogeneity of ISM in each subhalo as well as a merging process of subhalos during the evolutionary time of the Galactic halo (2 Gyr). The contribution of $s$-process elements also was excluded, which could be important for the evolutions of Ba and Sr at [Fe/H] $\gtrsim -2.5$. Nevertheless, our simplified approach enabled us to disentangle the different sources of dispersion: the subhalo-mass-dependent SFRs that lead to a spread in [Fe/H] for the same [$r$/Fe] (as shown by IWP15) and the stochastic NSM events that lead to a spread in [$r$/Fe] for the same [Fe/H]. However, our models failed to explain the moderate dispersion of the observed [$r$/Fe] ratios at [Fe/H] $\sim -1.5$ (Figure~4), which may be due to the combined effects of ISM inhomogeneity and $s$-process contribution. Such effects will be explored in our forthcoming paper.}


\acknowledgements
We thank Y Hirai for useful discussions. This work was supported by JSPS and CNRS under the Japan-France Research Cooperative Program (CNRS PRC No. 1363), the JSPS Grants-in-Aid for Scientific Research (26400232, 26400237), and the RIKEN iTHEMS Project.

\end{document}